\RequirePackage{lineno}
\documentclass[twocolumn,aps,prl,showpacs,superscriptaddress,floatfix]{revtex4}
\usepackage{epsfig}
\newcommand {\snn}	{\sqrt{s_{_{\rm NN}}}}
\newcommand {\gevc}	{GeV/$c$}
\newcommand {\dAu}	{$d$+Au}

\newcommand {\Ncoll}	{N_{\rm coll}}
\newcommand {\pt}	{p_{\perp}}
\newcommand {\rt}	{r_{\perp}}
\newcommand {\bt}	{\beta_{\perp}}
\newcommand {\hpt}	{\hat{p}_{\perp}}
\newcommand {\hrt}	{\hat{r}_{\perp}}
\newcommand {\mean}[1]	{\langle #1\rangle}
\newcommand {\rtpt}	{\mean{\hrt\cdot\hpt}}

\begin{document}
\title{Anisotropic parton escape is the dominant source of azimuthal anisotropy in transport models}
\author{Liang He}
\address{Department of Physics and Astronomy, Purdue University, West Lafayette, IN 47907, USA}
\author{Terrence Edmonds}
\address{Department of Physics, University of Florida, Gainesville, FL 32611, USA}
\author{Zi-Wei Lin}
\address{Department of Physics, East Carolina University, Greenville, NC 27858, USA}
\author{Feng Liu}
\address{Key Laboratory of Quark \& Lepton Physics (Central China Normal University), Ministry of Education, China}
\author{Denes Molnar}
\address{Department of Physics and Astronomy, Purdue University, West Lafayette, IN 47907, USA}
\author{Fuqiang Wang}
\email{fqwang@purdue.edu}
\address{Department of Physics and Astronomy, Purdue University, West Lafayette, IN 47907, USA}
\address{Key Laboratory of Quark \& Lepton Physics (Central China Normal University), Ministry of Education, China}

\begin{abstract}
We trace the development of azimuthal anisotropy ($v_n, n=2,3$) via parton-parton collision history in two transport models. The parton $v_n$ is studied as a function of the number of collisions of each parton in Au+Au and \dAu\ collisions at $\snn=200$~GeV. It is found that the majority of $v_n$ comes from the anisotropic escape probability of partons, with no fundamental difference at low and high transverse momenta. The contribution to $v_n$ from hydrodynamic-type collective flow is found to be small. Only when the parton-parton cross-section is set unrealistically large does this contribution start to take over. Our findings challenge the current paradigm emerged from hydrodynamic comparisons to anisotropy data.
\end{abstract}
\maketitle


Relativistic heavy ion collisions aim to create the quark-gluon plasma (QGP) and study quantum chromodynamics (QCD) at the extreme conditions of high temperature and energy density~\cite{Arsene:2004fa}. 
The system created in these collisions is described well by hydrodynamics where the high pressure buildup drives the system to expand at relativistic speed~\cite{Heinz:2013th}. 
Experimental data fit with hydrodynamics inspired models suggest that particles are locally thermalized and possess a common radial flow velocity~\cite{Abelev:2008ab}. Of particular interest are non-central collisions where the overlap volume of the colliding nuclei is anisotropic in the transverse plane (perpendicular to beam). The pressure gradient would generate anisotropic expansion and final-state elliptic flow~\cite{Ollitrault:1992bk}. Large elliptic anisotropy in momentum ($v_2$) has been measured, as large as hydrodynamic calculations predict~\cite{Arsene:2004fa,Heinz:2013th}. 
This suggests that the collision system is strongly interacting and nearly thermalized (sQGP)~\cite{Gyulassy:2004zy}. 

Molnar's Parton Cascade (MPC)~\cite{Molnar:2000jh} can describe the measured large $v_2$, but with an unusually large parton-parton interaction cross-section ($\sigma$)~\cite{Molnar:2001ux}. It approaches~\cite{Molnar:2004yh} the limiting case of ideal hydrodynamics ($\sigma\to\infty$) and may be an effective description of the sQGP. A Multi-Phase Transport (AMPT)~\cite{Zhang:1999bd,Lin:2004en} can describe the large anisotropy with $\sigma$ motivated by perturbative QCD but with the string melting mechanism~\cite{Lin:2001zk}. String melting liberates strings into a larger number of quarks and antiquarks, effectively increasing the system opacity~\cite{Lin:2014tya}.

It is generally perceived that large $v_2$ can only be generated in large-system heavy ion collisions. Recent particle correlation data, however, hint at similar $v_2$ in small systems of high multiplicity 
$p$+$p$ and $p$+Pb collisions at the LHC~\cite{Khachatryan:2010gv} 
and \dAu\ collisions at RHIC~\cite{Adare:2014keg}. 
Hydrodynamics has been applied to these systems and seems to describe the experimental data well~\cite{Bozek:2010pb}. 
AMPT also appears to describe the measured correlations~\cite{Bzdak:2014dia}. 
This suggests that these small-system collisions might create an sQGP as well, in contrast to the general expectations. 

The purpose of this Letter is to study the development of azimuthal anisotropy to shed light on its connection to the properties of the sQGP and thermalization. We employ the string melting version of AMPT~\cite{Lin:2004en} because it reasonably reproduces particle yields, transverse momentum ($\pt$) spectra, and $v_2$ data for the bulk (see Figs.~1-3 of Ref.~\cite{Lin:2014tya}). 
The model consists of a fluctuating initial condition, parton elastic scatterings, quark coalescence for hadronization, and hadronic interactions. In particular, parton scatterings are treated with Zhang's Parton Cascade (ZPC)~\cite{Zhang:1997ej}. We use Debye screened differential cross-section $d\sigma/dt\propto\alpha_s^2/(t-\mu_D^2)^2$~\cite{Lin:2004en} in AMPT, with strong coupling constant $\alpha_s=0.33$ and Debye screening mass $\mu_D=2.265$/fm (the total cross section is then $\sigma=3$~mb). We also employ MPC~\cite{Molnar:2000jh}, in order to check the generality and model dependence of our study. MPC employs the parton subdivision technique to eliminate acausal numerical artifacts due to action at a distance in the cascade algorithm. 
Here MPC is used with smooth, longitudinal boost invariant, binary collision profile for the initial conditions, as in Ref.~\cite{Molnar:2004yh}. We use both the Debye and isotropic $d\sigma/dt$ in MPC; the results are qualitatively similar. We analyze the entire history of parton-parton interactions in these models. For simplicity only partons are analyzed. 

We compute the $n^{\rm th}$ harmonic plane (short-axis direction of the corresponding harmonic component) of each event from its initial configuration of partons~\cite{Ollitrault:1993ba} by
\begin{equation}
\psi_n^{(r)}=\frac{1}{n}\left[{\rm atan2}(\mean{\rt^{2}\sin n\phi_r},\mean{\rt^{2}\cos n\phi_r})+\pi\right]\,.
\end{equation}
Here $\rt$ and $\phi_r$ are the polar coordinate of each initial parton (after its formation time) in the transverse plane, and $\mean{...}$ denotes the per-event average. We analyze the momentum anisotropy in the initial state, final state, and any intermediate state in-between. The momentum anisotropy is characterized by Fourier coefficients~\cite{Voloshin:1994mz}
\begin{equation}
v_n^{\rm obs}=\mean{\cos n(\phi -\psi_n^{(r)})}\,,
\end{equation}
where $\phi$ is the azimuthal angle of the parton momentum.

\begin{figure}[]
  \begin{center}
    \includegraphics[width=\columnwidth]{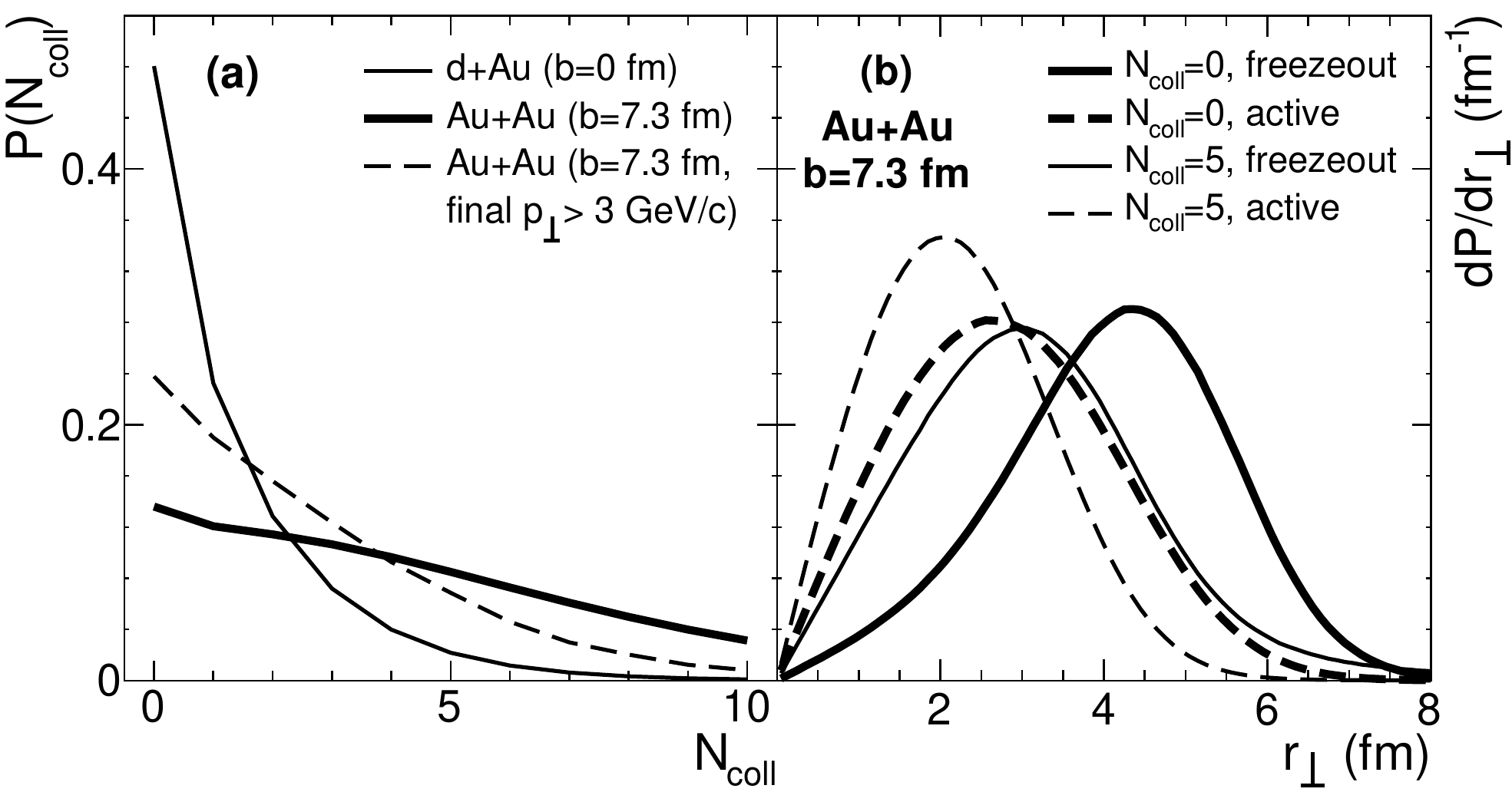}
    \caption{(a) Normalized probability distributions of partons freezing out after $\Ncoll$ collisions in AMPT (Debye $\sigma=3$~mb). The thin and thick solid curves are for partons of all $\pt$ in \dAu\ and Au+Au collisions, respectively. The dashed curve is for partons with final $\pt$$>$3~\gevc\ in Au+Au collisions. (b) Normalized probability distributions of parton transverse radius $\rt$ in Au+Au collisions from AMPT. The thick curves are for $\Ncoll$=0 and the thin curves for $\Ncoll$=5. The solid curves are for freezeout partons and the dashed curves for active partons.}
    \label{fig:ncoll_AMPT}
  \end{center}
\end{figure}

{\em Results.} 
We simulate Au+Au (impact parameter $b=7.3$~fm) and \dAu\ ($b=0$~fm) collisions by AMPT with $\sigma=3$~mb. We trace the history of parton cascading by the number of collisions ($\Ncoll$) a parton suffers with other partons. Figure~\ref{fig:ncoll_AMPT}(a) shows the probability distributions of partons freezing out after $\Ncoll$ collisions. Partons in mid-central Au+Au suffer more collisions than in d+Au, as expected. See Table~\ref{tab} for the average number of collisions (opacity), $\mean{\Ncoll}$, for each collision system. As seen in Fig.~\ref{fig:ncoll_AMPT}(a), partons with higher final $\pt$ have fewer collisions and freeze out earlier. 

As shown in Fig.~\ref{fig:ncoll_AMPT}(a), some partons do not interact at all and thus instantly freeze out at $\Ncoll$=0. These partons tend to reside in the outer region of the overlap volume (``surface emission''), as shown in Fig.~\ref{fig:ncoll_AMPT}(b) for Au+Au where the transverse radius ($\rt$) distribution of freezeout partons is depicted by the thick solid curve. Those continuing to interact tend to be inside as shown by the thick dashed curve. This feature is qualitatively similar for all $\Ncoll$ values (e.g.~see the thin curves in Fig.~\ref{fig:ncoll_AMPT}(b) for $\Ncoll$=5). This is consistent with the general expectation--the energy density is smaller in the outer shell thus the probability for further interactions is smaller. It is interesting to note that the freezeout ``surface'' moves inward, indicating an outside-to-inside freezeout scenario.

\begin{figure}[]
  \begin{center}
    \includegraphics[width=\columnwidth]{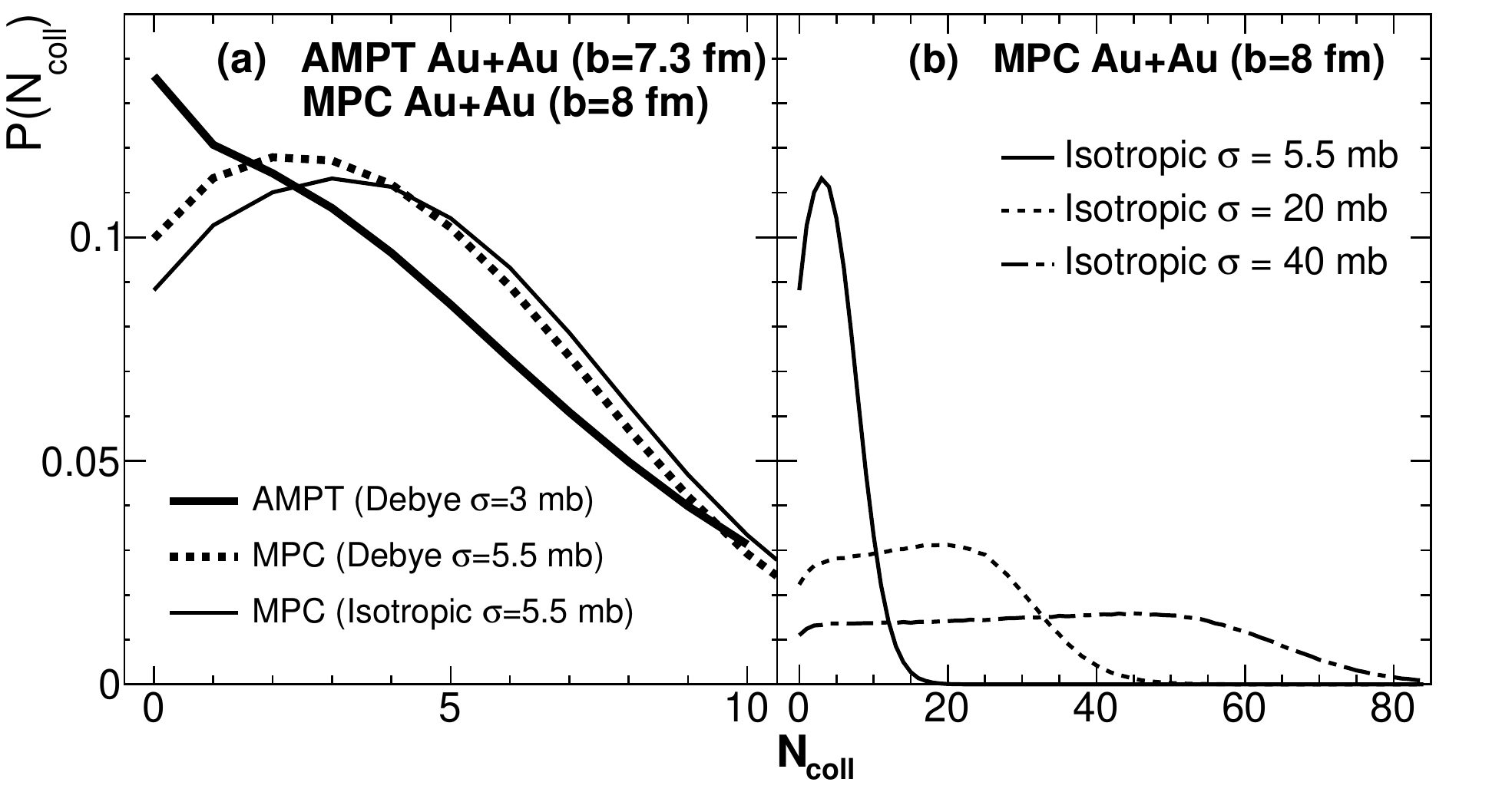}
    \caption{(a) Comparison of the normalized probability distributions of freezeout partons in Au+Au collisions from AMPT (Debye $\sigma=3$~mb) and MPC (both Debye and isotropic $\sigma=5.5$~mb) of comparable opacities. (b) Normalized probability distributions of partons freezing out after $\Ncoll$ collisions in MPC with three isotropic $\sigma$ values: 5.5~mb (solid), 20~mb (dashed), and 40~mb (dash-dotted).}
    \label{fig:ncoll_MPC}
  \end{center}
\end{figure}

Due to different initial parton densities, we use Debye $\sigma=5.5$~mb for MPC to obtain a similar opacity $\mean{\Ncoll}$=4-5 as AMPT with Debye $\sigma=3$~mb. The $\Ncoll$ distributions are similar between AMPT and MPC, as illustrated in Fig.~\ref{fig:ncoll_MPC}(a). Also shown is the MPC $\Ncoll$ distribution with isotropic $\sigma=5.5$~mb which is again similar. Figure~\ref{fig:ncoll_MPC}(b) shows the $\Ncoll$ distributions from MPC with three isotropic $\sigma$ values: 5.5~mb, 20~mb, and 40~mb. As $\sigma$ increases, $\mean{\Ncoll}$ becomes larger as expected. The probability for small $\Ncoll$ is, nevertheless, non-zero even at large $\sigma$ because of finite surface emission. 

\begin{figure}{}
  \begin{center}
    \includegraphics[width=\columnwidth]{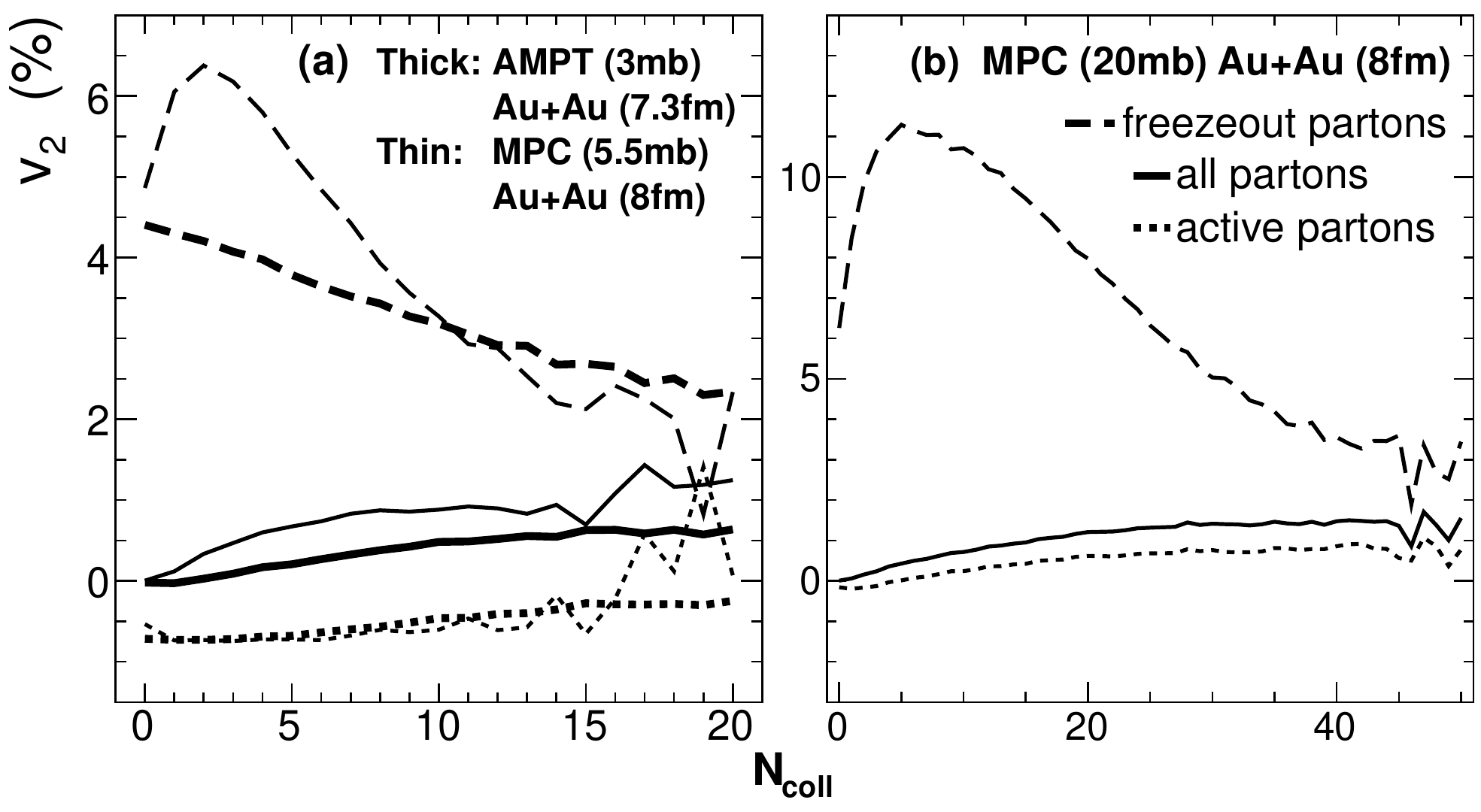}
    \caption{Parton $v_2$ in Au+Au collisions after suffering $\Ncoll$ collisions in (a) AMPT with Debye $\sigma=3$~mb (thick curves) and MPC with isotropic $\sigma=5.5$~mb (thin curves), and (b) MPC with isotropic $\sigma=20$~mb. The dashed curves are for partons freezing out after exactly $\Ncoll$ collisions, the dotted curves for partons continuing to interact, and the solid curves for all partons (i.e. sum of the former two).}
    \label{fig:v2}
  \end{center}
\end{figure}

We track the development of $v_2$ in AMPT and MPC by studying parton $v_2$ as a function of $\Ncoll$ in Fig.~\ref{fig:v2} for Au+Au collisions. The solid curves are the $v_2$ of all partons after suffering $\Ncoll$ collisions (those frozen out with smaller $\Ncoll$ values are of course not included). The dashed curves are the $v_2$ of the partons that freeze out after suffering exactly $\Ncoll$ collisions (i.e.~without further interactions). At $\Ncoll=0$, the $v_2$ of all partons is zero because parton form with axially uniform momenta. Some partons (``corona'') do not interact at all and instantly freeze out with $\Ncoll=0$. Because there is a larger probability for the partons to escape along the short axis of the overlap volume, those freezeout partons have positive $v_2$~\cite{Molnar:2005hb,Lin:2014tya}. 
In the low density limit (LDL), the anisotropy may be analytically derived~\cite{Heiselberg:1998es,Kolb:2000fha}. 
In fact, this escape mechanism is rather general as it happens throughout the entire evolution of the collision system. After $\Ncoll$ collisions, the $v_2$ of all partons is still roughly zero. Some of these partons freeze out; they have positive $v_2$ partly due to the preferential escape along the short axis. The remaining partons, that can have negative $v_2$ as shown by the dotted curves in Fig.~\ref{fig:v2}(a), continue to interact. With one more collision, the azimuthal distribution of those partons becomes roughly isotropic again, with approximately zero $v_2$ (solid curves). This process then repeats itself. 

The $v_2$ results are similar between AMPT with Debye $\sigma=3$~mb and MPC with Debye $\sigma=5.5$~mb, both corresponding to a similar opacity $\mean{\Ncoll}$ (see Table~\ref{tab}). 
In Fig.~\ref{fig:v2}(a) we show instead the MPC results with isotropic $\sigma=5.5$~mb to compare to the AMPT results. The results are qualitatively similar while the freezeout $\mean{v_2}$ is larger in MPC. MPC results with a larger opacity are shown in Fig.~\ref{fig:v2}(b). The escape picture still holds at large opacities. However, the probabilities to escape at small $\Ncoll$ are now smaller, so the escape contribution to the final overall $v_2$ is smaller, and the remaining active partons have mostly positive $v_2$. 

\begin{figure}{}
  \begin{center}
    \includegraphics[width=0.7\columnwidth]{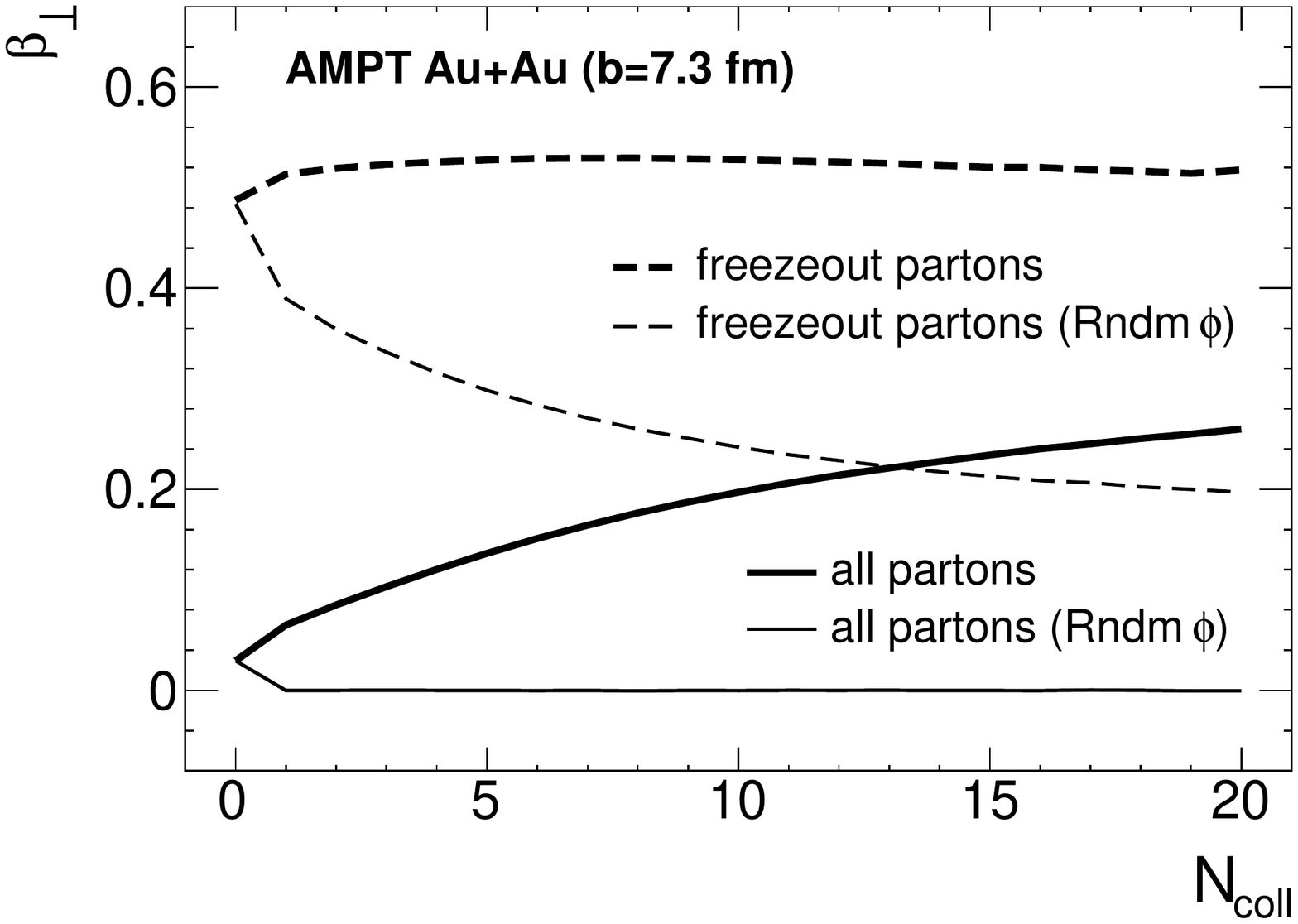}
    \caption{Parton $\bt\equiv\rtpt$ as a function of $\Ncoll$ in Au+Au collisions. Both normal (thick curves) and azimuth-randomized (thin curves) AMPT results are shown. The solid curves are those for all partons after suffering $\Ncoll$ collisions, and the dashed curves are for freezeout partons (partons that freeze out after suffering exactly $\Ncoll$ collisions).}
    \label{fig:rp}
  \end{center}
\end{figure}

Figure~\ref{fig:rp} shows the approximate average transverse radial velocity of partons in AMPT, $\bt\equiv\rtpt$ where $\hrt$ and $\hpt$ are the transverse radial position and momentum unit vectors, as a function of $\Ncoll$. The $\bt$ of all partons (thick solid curve) at $\Ncoll=0$ is not exactly zero because partons can form only after a finite formation time over which a parton's displacement depends on its momentum. The freezeout partons (thick dashed curve) at $\Ncoll=0$ have a large $\bt$. This strong space-momentum correlation is due to the anisotropic escape mechanism, but different from a collectivity that represents a common collective {\em flow} velocity achieved only via interactions. On the other hand, there are space-momentum correlations for all partons (thick solid curve) at any given $\Ncoll>0$ and the correlation increases with $\Ncoll$; these correlations are good indicators of collective flow. Some of these partons freeze out at a given $\Ncoll$; the additional $\bt$ for these freezeout partons is the effect of the anisotropic escape mechanism. 

One question is whether the hydrodynamic-type collective flow of the partons is important for the final $v_2$. Thus we did test calculations with no collective flow by randomizing the outgoing parton azimuthal directions after each parton-parton scattering. The system continues to evolve in AMPT, but the evolution is different from the original one. The $\bt$ from this modified evolution is shown in Fig.~\ref{fig:rp}. The all-parton $\bt$ is now zero because of the randomization, and the freezeout parton $\bt$ is non-zero purely due to the anisotropic escape mechanism. 

\begin{figure}{}
  \begin{center}
    \includegraphics[width=\columnwidth]{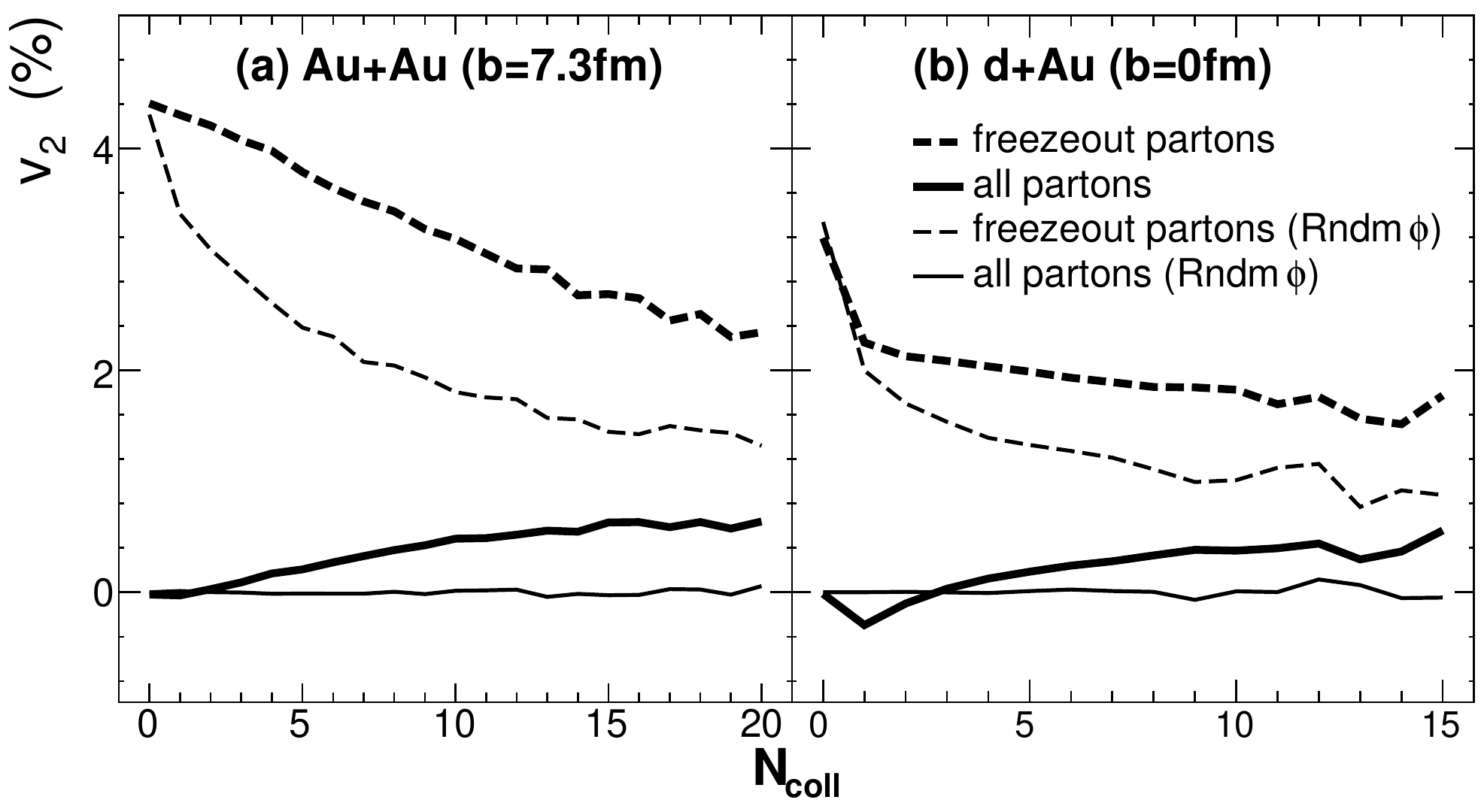}
    \caption{Parton $v_2$ as a function of $\Ncoll$ in (a) Au+Au and (b) \dAu\ collisions. Both normal (thick curves) and azimuth-randomized (thin curves) AMPT results are shown. The solid curves are for all partons after suffering $\Ncoll$ collisions, and the dashed curves are for freezeout partons.}
    \label{fig:v2_rndm}
  \end{center}
\end{figure}

We show in Fig.~\ref{fig:v2_rndm} the $v_2$ of all partons and freezeout partons from this azimuth-randomized AMPT by the thin solid and dashed curves for Au+Au and \dAu\ collisions. In the randomized case, the parton azimuthal angles are randomized and hence their $v_2$ is zero; thus the final-state freezeout anisotropy is entirely due to the anisotropic escape mechanism. For comparison, the $v_2$ results from the normal AMPT (already shown in Fig.~\ref{fig:v2} for Au+Au) are superimposed in Fig.~\ref{fig:v2_rndm} as the thick solid and dashed curves, where the all-parton $v_2$ is slightly positive and the freezeout parton $v_2$ is much higher. The gain in $v_2$ by the freezeout partons is due to the escape mechanism. The gain in the normal AMPT results is slightly different from that in the azimuth-randomized results. This is not surprising because the anisotropies in the escape probability differ in these two cases: in the former case the parton $\hpt$'s are correlated with their $\hrt$'s while in the latter case the parton $\hpt$'s are random. 

\begin{figure}{}
  \begin{center}
    \includegraphics[width=0.6\columnwidth]{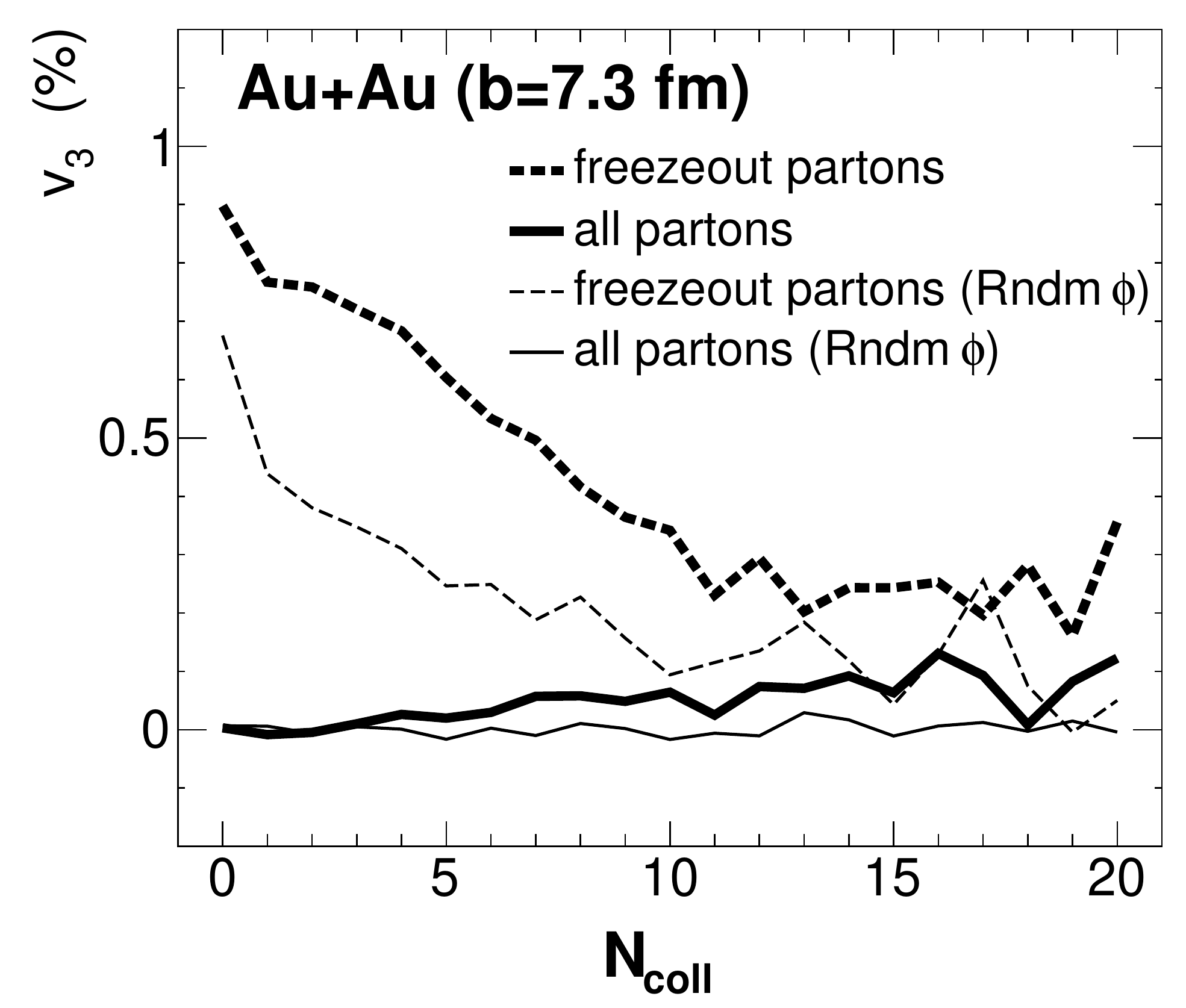}
    \caption{Parton $v_3$ as a function of $\Ncoll$ in Au+Au collisions. Both normal (thick curves) and azimuth-randomized (thin curves) AMPT results are shown. The solid curves are for all partons after suffering $\Ncoll$ collisions, and the dashed curves are for freezeout partons.}
    \label{fig:v3_rndm}
  \end{center}
\end{figure}

We have shown mostly results for 200 GeV Au+Au collisions at medium impact parameters. There seems to be no qualitative difference between the behaviors in Au+Au and \dAu\ collisions (c.f.~Fig.~\ref{fig:v2_rndm}). Although we focused on $v_2$, the same qualitative conclusions hold for $v_3$ as well--see Fig.~\ref{fig:v3_rndm} where $v_3$ is shown similar to Fig.~\ref{fig:v2_rndm}(a). This suggests that the development mechanism of anisotropies in transport models is universal. We note that a fixed $\Ncoll$ value does not correspond to partons at identical time but rather a convolution over time. We have also studied results as a function of time instead of $\Ncoll$, and our qualitative conclusions remain unchanged. 

\begin{figure}{}
  \begin{center}
    \includegraphics[width=\columnwidth]{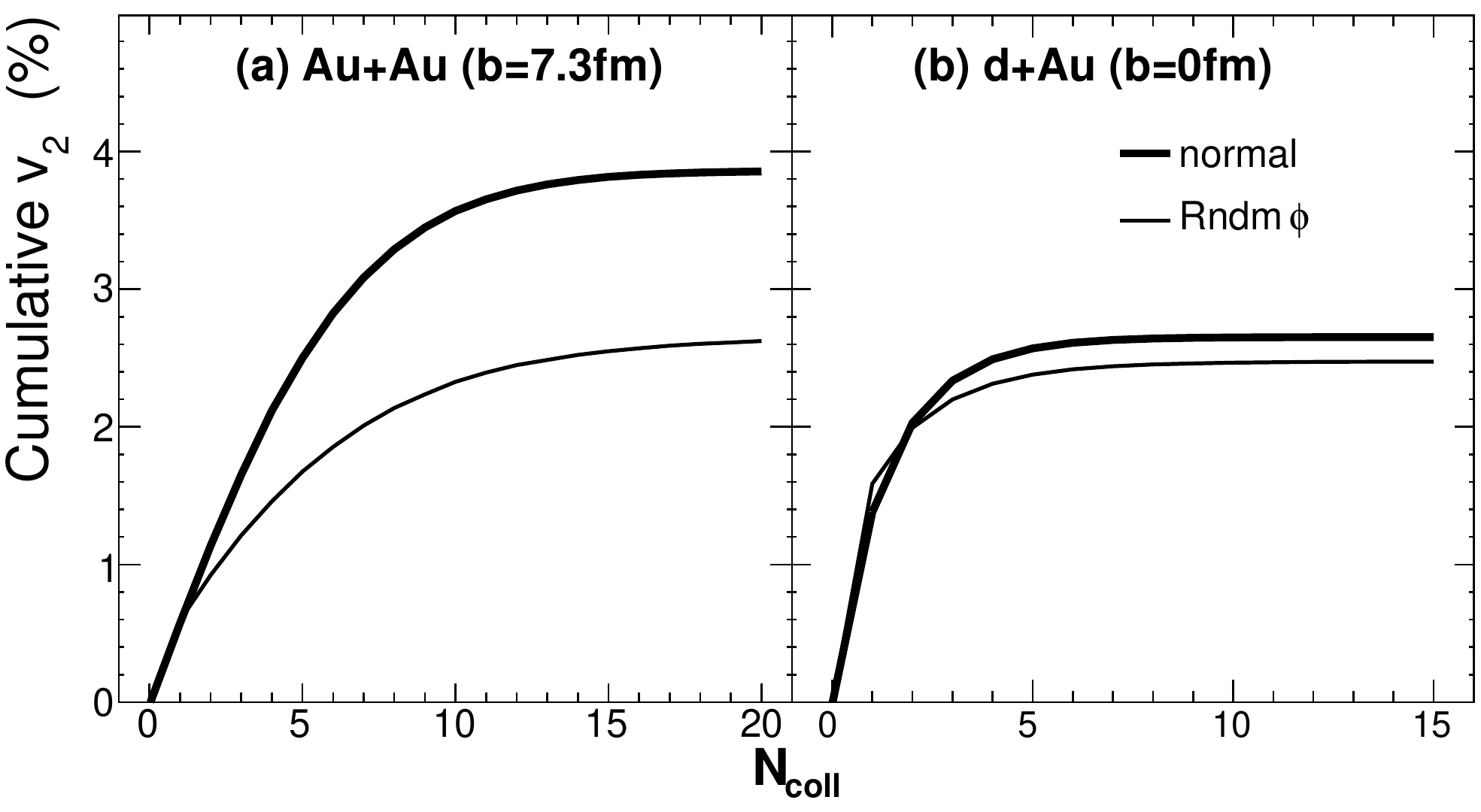}
    \caption{Cumulative $v_2$ of all partons 
(see text) as a function of $\Ncoll$ in (a) Au+Au and (b) \dAu\ collisions. Both normal (thick curves) and azimuth-randomized (thin curves) AMPT results are shown.}
    \label{fig:v2cumu}
  \end{center}
\end{figure}

Experimentally, only the final-state anisotropy is measured, integrated over all evolution time. 
It is thus interesting to examine the cumulative $v_2$ of all partons up to $\Ncoll$ collisions, including those that have frozen out with $\Ncoll$ or fewer collisions and those that will suffer further collisions, as a function of $\Ncoll$. This is shown in Fig.~\ref{fig:v2cumu}. The cumulative $v_2$ increases with $\Ncoll$ and starts to saturate after approximately 10 and 3 collisions in mid-central Au+Au and central d+Au collisions, respectively. The asymptotic $v_2$ value at $\Ncoll\to\infty$ would be the final-state parton average $\mean{v_2}$ in the events.

Table~\ref{tab} lists the $\mean{\Ncoll}$ and $\mean{v_2}$ of all final partons from normal and azimuth-randomized results by both AMPT (varying $\mu_D$ with $\alpha_s$ kept fixed) and MPC (isotropic $\sigma$). The $\mean{v_2}$ with Debye $\sigma$ of 3~mb and 5.5~mb from AMPT and MPC, respectively--so that they have similar opacity--are consistent with each other as mentioned previously. It is known that $v_2$ is larger for an isotropic $d\sigma/dt$ than for a Debye screened one (at the same $\sigma$) because the latter is more forward-backward peaked, but the $\mean{v_2}_{\rm Rndm}/\mean{v_2}$ ratios as well as the opacities are almost the same. It should be noted that there may be issues with causality in AMPT for large $\sigma$. Also note that $\mean{\Ncoll}$ is larger in the randomized case because the randomization tends to destroy the preferred outward direction of partons. 
\begin{table*}
\caption{$\mean{\Ncoll}$ and $\mean{v_2}$ of all final partons from normal (first number in each column) and azimuth-randomized AMPT (Debye $\sigma$) and MPC (isotropic $\sigma$) results (second number). The \dAu\ impact parameter is $b=0$~fm.}
\label{tab}
\begin{tabular}{c|cc|cc|cc|cc|cc|cc|cc|cc|cc}\hline
& \multicolumn{2}{c|}{AMPT \dAu} & \multicolumn{8}{c|}{AMPT Au+Au ($b=7.3$~fm)} & \multicolumn{8}{c}{MPC Au+Au ($b=8$~fm)} \\\hline
$\sigma$ & \multicolumn{2}{c|}{3~mb} & \multicolumn{2}{c|}{3~mb} & \multicolumn{2}{c|}{20~mb} & \multicolumn{2}{c|}{40~mb} & \multicolumn{2}{c|}{60~mb} & \multicolumn{2}{c|}{5.5~mb} & \multicolumn{2}{c|}{20~mb} & \multicolumn{2}{c|}{40~mb} & \multicolumn{2}{c}{60~mb} \\\hline
$\mean{\Ncoll}$ & 1.2 & 1.4 & 4.6 & 5.8 & 13 & 22 & 17 & 32 & 20 & 39 & 4.7 & 5.4 & 17 & 23 & 35 & 52 & 53 & 83 \\
$\mean{v_2}$ & 2.7\% & 2.5\% & 3.9\% & 2.7\% & 5.9\% & 2.7\% & 6.0\% & 2.3\% & 5.7\% & 2.0\% & 5.5\% & 3.5\% & 8.6\% & 3.5\% & 9.8\% & 3.0\% & 10\% & 2.6\% \\\hline
$\mean{v_2}_{\rm Rndm}/\mean{v_2}$ & \multicolumn{2}{c|}{93\%} & \multicolumn{2}{c|}{69\%} & \multicolumn{2}{c|}{46\%} & \multicolumn{2}{c|}{38\%} & \multicolumn{2}{c|}{34\%} & \multicolumn{2}{c|}{64\%} & \multicolumn{2}{c|}{41\%} & \multicolumn{2}{c|}{31\%} & \multicolumn{2}{c}{26\%} \\\hline
\end{tabular}
\end{table*}

{\em Discussions.}
The unique finding of our study is that partons in transport models escape (freeze out) from the collision zone with positive $v_2$, even those partons that do not interact at all, mainly due to the anisotropic escape probability. This escape mechanism contributes to the majority of the final $v_2$ at small to modest opacity. The traditional picture of low $\pt$ particles accumulating $v_2$ after multiple collisions seems to play only a minor role. High-$\pt$ anisotropy is generally believed to result mostly from the escape mechanism~\cite{Wang:1991xy,Shuryak:2001me}, which we have also verified within our work. Our study indicates that the escape mechanism is at work at both high and low $\pt$; there appears to be no fundamental difference in the $v_2$ development of high- and low-$\pt$ partons. The $\pt$ dependence of $v_2$--also captured in AMPT--is less obvious from the escape mechanism. It is partly borne out of the fact that higher-$\pt$ partons freeze out after fewer collisions (c.f.~Fig.~\ref{fig:ncoll_AMPT}(a)), and hence possess larger $v_2$ (c.f.~Fig.~\ref{fig:v2}). The LDL calculation of Ref.~\cite{Kolb:2000fha} found that the centrality dependence and the magnitude of $\pt$-averaged elliptic flow were well described, but the shape of the $\pt$ dependence was siginificantly off at RHIC. 

It has generally been thought that the reason why AMPT describes the bulk experimental data well~\cite{Lin:2014tya} is because AMPT has large opacity and thus has approached hydrodynamics. Our study indicates that this interpretation is incorrect and in fact the opacity in AMPT is only small to modest. If one takes the $\mean{v_2}_{\rm Rndm}$ from the azimuth-randomized results as estimate of the escape contribution to the $\mean{v_2}$ from the normal results, then its contribution in semi-central Au+Au collisions is $\sim$70\% with modest opacity $\mean{\Ncoll}$=4-5. As opacity increases, the hydrodynamic-type collective flow contribution increases, but rather slowly. The system is still far away from asymptotic hydrodynamic behavior even with the unrealistically large opacities we have studied. It is found that this picture is qualitatively the same for $v_3$. 

The space-momentum correlations of freezeout partons in the transport models are largely due to the escape mechanism. This results in ``surface emission,''~\cite{Shuryak:2001me} where a parton freezes out depending on its momentum and position, which determine its escape probability at that point of evolution. 
It thus creates a space-momentum correlation even in the absence of hydrodynamic collective flow, as demonstrated by the azimuth-randomized AMPT results. 
It is important here to distinguish between space-momentum correlation and hydrodynamic collective flow, where the latter means a collective motion that is generated by interactions so that particles convert part of their energy into their common motion (e.g.~particles in nearly local thermal equilibrium moving on top of a common velocity field). There is indeed a finite collective flow in AMPT (the thick solid curve in Fig.~\ref{fig:rp}). This radial flow is presumably generated by hydrodynamic-type interactions and pressure gradient. The azimuthal modulation of this radial flow is the anisotropic flow $v_n$ of partons (the solid thick curves in Figs.~\ref{fig:v2},~\ref{fig:v2_rndm}, and~\ref{fig:v3_rndm}). It is the $v_n$ of these active partons that is the most relevant for the standard hydrodynamic flow description, or the collective properties of the sQGP. The radial flow may be a viable discriminator: collectivity generates extra $\pt$ but the escape mechanism does not. Experimentally, however, radial flow may be difficult to measure as it requires precise knowledge of initial-state transverse energy production. The centrality dependence of $v_n$ may serve as another possible discriminator, since hydrodynamics and LDL give different predictions~\cite{Heiselberg:1998es,Kolb:2000fha}.

Hydrodynamics has been successfully applied to heavy ion collisions~\cite{Heinz:2013th}, and a small viscosity to entropy density ratio ($\eta/s$), close to the conjectured quantum limit of $1/4\pi$, has been extracted~\cite{Kovtun:2004de}. 
Hydrodynamic evolution is typically stopped in a calculation when the {\em local} energy density or temperature reaches a given value. Particle production is then modeled by the Cooper-Frye formalism~\cite{Cooper:1974mv}. The escape mechanism, on the other hand, is driven by the chance of no further interaction $P_{\rm escape}=\exp\left(-\int\rho\sigma d\ell\right)$, i.e., a non-local quantity that involves the entire future density ($\rho$) evolution of the system, and is not obviously captured by the Cooper-Frye prescription. If hydrodynamically driven collective flow is indeed a small contribution to the experimentally measured anisotropy, then the extracted $\eta/s$ is severely underestimated. Since escape is inevitable for our transient collision systems, it is imperative to examine the possible role of the escape mechanism in the hydrodynamics framework. Previous studies have shown that continuous particle emission instead of sudden freezeout in hydrodynamics can have important implications for pion interferometry~\cite{Grassi:1994nf}. 

In summary, we have studied the development of azimuthal anisotropy $v_n (n=2,3)$ in AMPT and MPC as a function of the number of collisions $\Ncoll$ that a parton suffers in Au+Au and \dAu\ collisions at $\snn=200$~GeV. It is found that the majority of $v_n$ comes from the anisotropic escape probability of partons, and this picture applies similarly to partons at both high $\pt$ and low $\pt$. The anisotropic flow of partons as a result of parton-parton interactions or hydrodynamic-type pressure gradient is found to be small in transport models. This part of anisotropy becomes more important with increasing cross-section $\sigma$, but the system in transport models is still far from the asymptotic hydrodynamic behavior even with the unrealistically large cross-sections that we studied. The escape mechanism is dominant because the evolution in transport models is relatively dilute ($\mean{\Ncoll}$=4-5 at $\sigma\sim3$~mb). This is distinctly different from evolution near the hydrodynamic limit, where collectivity is generated by a large number of collisions. Such hydrodynamic-type collectivity and anisotropic flow is what one would regard as the cornerstone of the sQGP paradigm. Our results, however, suggest that the assumption of hydrodynamics is not imperative.

This work is supported in part by US~Department of Energy Grant No.~DE-FG02-88ER40412 (HL,FW) and No.~DE-FG02-13ER16413 (DM), the National Science Foundation Research Experience for Undergraduates Grant No.~1156764 (TE), and National Natural Science Foundation of China Grant No.~11228513 (FL).


\begin{thebibliography}{26}
\expandafter\ifx\csname natexlab\endcsname\relax\def\natexlab#1{#1}\fi
\expandafter\ifx\csname bibnamefont\endcsname\relax
  \def\bibnamefont#1{#1}\fi
\expandafter\ifx\csname bibfnamefont\endcsname\relax
  \def\bibfnamefont#1{#1}\fi
\expandafter\ifx\csname citenamefont\endcsname\relax
  \def\citenamefont#1{#1}\fi
\expandafter\ifx\csname url\endcsname\relax
  \def\url#1{\texttt{#1}}\fi
\expandafter\ifx\csname urlprefix\endcsname\relax\def\urlprefix{URL }\fi
\providecommand{\bibinfo}[2]{#2}
\providecommand{\eprint}[2][]{\url{#2}}

\bibitem[{\citenamefont{Arsene et~al.}(2005)}]{Arsene:2004fa}
\bibinfo{author}{\bibfnamefont{I.}~\bibnamefont{Arsene}} \bibnamefont{et~al.}
  (\bibinfo{collaboration}{BRAHMS Collaboration}),
  \bibinfo{journal}{Nucl.Phys.} \textbf{\bibinfo{volume}{A757}},
  \bibinfo{pages}{1} (\bibinfo{year}{2005}); 
%
\bibinfo{author}{\bibfnamefont{B.}~\bibnamefont{Back}} \bibnamefont{et~al.}
  (\bibinfo{collaboration}{PHOBOS Collaboration}),
  \bibinfo{journal}{{\it ibid.},} 
  \bibinfo{pages}{28}; 
%
\bibinfo{author}{\bibfnamefont{J.}~\bibnamefont{Adams}} \bibnamefont{et~al.}
  (\bibinfo{collaboration}{STAR Collaboration}), 
  \bibinfo{journal}{{\it ibid.},} 
  \bibinfo{pages}{102}; 
%
\bibinfo{author}{\bibfnamefont{K.}~\bibnamefont{Adcox}} \bibnamefont{et~al.}
  (\bibinfo{collaboration}{PHENIX Collaboration}),
  \bibinfo{journal}{{\it ibid.},} 
  \bibinfo{pages}{184}; 
%
\bibinfo{author}{\bibfnamefont{B.}~\bibnamefont{Muller}},
  \bibinfo{author}{\bibfnamefont{J.}~\bibnamefont{Schukraft}},
  \bibnamefont{and} \bibinfo{author}{\bibfnamefont{B.}~\bibnamefont{Wyslouch}},
  \bibinfo{journal}{Ann.Rev.Nucl.Part.Sci.} \textbf{\bibinfo{volume}{62}},
  \bibinfo{pages}{361} (\bibinfo{year}{2012}). 

\bibitem[{\citenamefont{Heinz and Snellings}(2013)}]{Heinz:2013th}
\bibinfo{author}{\bibfnamefont{U.}~\bibnamefont{Heinz}} \bibnamefont{and}
  \bibinfo{author}{\bibfnamefont{R.}~\bibnamefont{Snellings}},
  \bibinfo{journal}{Ann.Rev.Nucl.Part.Sci.} \textbf{\bibinfo{volume}{63}},
  \bibinfo{pages}{123} (\bibinfo{year}{2013}); 
%
\bibinfo{author}{\bibfnamefont{C.}~\bibnamefont{Gale}},
  \bibinfo{author}{\bibfnamefont{S.}~\bibnamefont{Jeon}}, \bibnamefont{and}
  \bibinfo{author}{\bibfnamefont{B.}~\bibnamefont{Schenke}},
  \bibinfo{journal}{Int.J.Mod.Phys.} \textbf{\bibinfo{volume}{A28}},
  \bibinfo{pages}{1340011} (\bibinfo{year}{2013}). 

\bibitem[{\citenamefont{Abelev et~al.}(2009)}]{Abelev:2008ab}
\bibinfo{author}{\bibfnamefont{B.}~\bibnamefont{Abelev}} \bibnamefont{et~al.}
  (\bibinfo{collaboration}{STAR Collaboration}), \bibinfo{journal}{Phys.Rev.}
  \textbf{\bibinfo{volume}{C79}}, \bibinfo{pages}{034909}
  (\bibinfo{year}{2009}). 

\bibitem[{\citenamefont{Ollitrault}(1992)}]{Ollitrault:1992bk}
\bibinfo{author}{\bibfnamefont{J.-Y.} \bibnamefont{Ollitrault}},
  \bibinfo{journal}{Phys.Rev.} \textbf{\bibinfo{volume}{D46}},
  \bibinfo{pages}{229} (\bibinfo{year}{1992}).

\bibitem[{\citenamefont{Gyulassy and McLerran}(2005)}]{Gyulassy:2004zy}
\bibinfo{author}{\bibfnamefont{M.}~\bibnamefont{Gyulassy}} \bibnamefont{and}
  \bibinfo{author}{\bibfnamefont{L.}~\bibnamefont{McLerran}},
  \bibinfo{journal}{Nucl.Phys.} \textbf{\bibinfo{volume}{A750}},
  \bibinfo{pages}{30} (\bibinfo{year}{2005}). 

\bibitem[{\citenamefont{Molnar and Gyulassy}(2002)}]{Molnar:2001ux}
\bibinfo{author}{\bibfnamefont{D.}~\bibnamefont{Molnar}} \bibnamefont{and}
  \bibinfo{author}{\bibfnamefont{M.}~\bibnamefont{Gyulassy}},
  \bibinfo{journal}{Nucl.Phys.} \textbf{\bibinfo{volume}{A697}},
  \bibinfo{pages}{495} (\bibinfo{year}{2002}). 

\bibitem[{\citenamefont{Molnar and Huovinen}(2005)}]{Molnar:2004yh}
\bibinfo{author}{\bibfnamefont{D.}~\bibnamefont{Molnar}} \bibnamefont{and}
  \bibinfo{author}{\bibfnamefont{P.}~\bibnamefont{Huovinen}},
  \bibinfo{journal}{Phys.Rev.Lett.} \textbf{\bibinfo{volume}{94}},
  \bibinfo{pages}{012302} (\bibinfo{year}{2005}). 

\bibitem[{\citenamefont{Zhang et~al.}(2000)\citenamefont{Zhang, Ko, Li, and
  Lin}}]{Zhang:1999bd}
\bibinfo{author}{\bibfnamefont{B.}~\bibnamefont{Zhang}},
  \bibinfo{author}{\bibfnamefont{C.M.}~\bibnamefont{Ko}},
  \bibinfo{author}{\bibfnamefont{B.-A.} \bibnamefont{Li}}, \bibnamefont{and}
  \bibinfo{author}{\bibfnamefont{Z.-W.} \bibnamefont{Lin}},
  \bibinfo{journal}{Phys.Rev.} \textbf{\bibinfo{volume}{C61}},
  \bibinfo{pages}{067901} (\bibinfo{year}{2000}). 

\bibitem[{\citenamefont{Lin et~al.}(2005)\citenamefont{Lin, Ko, Li, Zhang, and
  Pal}}]{Lin:2004en}
\bibinfo{author}{\bibfnamefont{Z.-W.} \bibnamefont{Lin}},
  \bibinfo{author}{\bibfnamefont{C.~M.} \bibnamefont{Ko}},
  \bibinfo{author}{\bibfnamefont{B.-A.} \bibnamefont{Li}},
  \bibinfo{author}{\bibfnamefont{B.}~\bibnamefont{Zhang}}, \bibnamefont{and}
  \bibinfo{author}{\bibfnamefont{S.}~\bibnamefont{Pal}},
  \bibinfo{journal}{Phys.Rev.} \textbf{\bibinfo{volume}{C72}},
  \bibinfo{pages}{064901} (\bibinfo{year}{2005}). 

\bibitem[{\citenamefont{Lin and Ko}(2002)}]{Lin:2001zk}
\bibinfo{author}{\bibfnamefont{Z.-W.} \bibnamefont{Lin}} \bibnamefont{and}
  \bibinfo{author}{\bibfnamefont{C.M.}~\bibnamefont{Ko}},
  \bibinfo{journal}{Phys.Rev.} \textbf{\bibinfo{volume}{C65}},
  \bibinfo{pages}{034904} (\bibinfo{year}{2002}). 

\bibitem[{\citenamefont{Lin}(2014)}]{Lin:2014tya}
\bibinfo{author}{\bibfnamefont{Z.-W.} \bibnamefont{Lin}},
  \bibinfo{journal}{Phys.Rev.} \textbf{\bibinfo{volume}{C90}},
  \bibinfo{pages}{014904} (\bibinfo{year}{2014}). 

\bibitem[{\citenamefont{Khachatryan et~al.}(2010)}]{Khachatryan:2010gv}
\bibinfo{author}{\bibfnamefont{V.}~\bibnamefont{Khachatryan}}
  \bibnamefont{et~al.} (\bibinfo{collaboration}{CMS Collaboration}),
  \bibinfo{journal}{JHEP} \textbf{\bibinfo{volume}{1009}}, \bibinfo{pages}{091}
  (\bibinfo{year}{2010}); 
%
\bibinfo{author}{\bibfnamefont{S.}~\bibnamefont{Chatrchyan}}
  \bibnamefont{et~al.} (\bibinfo{collaboration}{CMS Collaboration}),
  \bibinfo{journal}{Phys.Lett.} \textbf{\bibinfo{volume}{B718}},
  \bibinfo{pages}{795} (\bibinfo{year}{2013}); 
%
\bibinfo{author}{\bibfnamefont{B.}~\bibnamefont{Abelev}} \bibnamefont{et~al.}
  (\bibinfo{collaboration}{ALICE Collaboration}), \bibinfo{journal}{{\it ibid.}} 
  \textbf{\bibinfo{volume}{B719}}, \bibinfo{pages}{29} (\bibinfo{year}{2013}); 
%
\bibinfo{author}{\bibfnamefont{G.}~\bibnamefont{Aad}} \bibnamefont{et~al.}
  (\bibinfo{collaboration}{ATLAS Collaboration}),
  \bibinfo{journal}{Phys.Rev.Lett.} \textbf{\bibinfo{volume}{110}},
  \bibinfo{pages}{182302} (\bibinfo{year}{2013}). 

\bibitem[{\citenamefont{Adare et~al.}(2014)}]{Adare:2014keg}
\bibinfo{author}{\bibfnamefont{A.}~\bibnamefont{Adare}} \bibnamefont{et~al.}
  (\bibinfo{collaboration}{PHENIX Collaboration}) (\bibinfo{year}{2014}), \eprint{1404.7461}.
%
\bibinfo{author}{\bibfnamefont{L.}~\bibnamefont{Adamczyk}} \bibnamefont{et~al.}
  (\bibinfo{collaboration}{STAR Collaboration}), \bibinfo{journal}{Phys.Lett.}
  \textbf{\bibinfo{volume}{B747}}, \bibinfo{pages}{265} (\bibinfo{year}{2015}). 

\bibitem[{\citenamefont{Bozek}(2011)}]{Bozek:2010pb}
\bibinfo{author}{\bibfnamefont{P.}~\bibnamefont{Bozek}},
  \bibinfo{journal}{Eur.Phys.J.} \textbf{\bibinfo{volume}{C71}},
  \bibinfo{pages}{1530} (\bibinfo{year}{2011}); 
%
\bibinfo{author}{\bibfnamefont{P.}~\bibnamefont{Bozek}} \bibnamefont{and}
  \bibinfo{author}{\bibfnamefont{W.}~\bibnamefont{Broniowski}},
  \bibinfo{journal}{Phys.Lett.} \textbf{\bibinfo{volume}{B718}},
  \bibinfo{pages}{1557} (\bibinfo{year}{2013}). 

\bibitem[{\citenamefont{Bzdak and Ma}(2014)}]{Bzdak:2014dia}
\bibinfo{author}{\bibfnamefont{A.}~\bibnamefont{Bzdak}} \bibnamefont{and}
  \bibinfo{author}{\bibfnamefont{G.-L.} \bibnamefont{Ma}},
  \bibinfo{journal}{Phys.Rev.Lett.} \textbf{\bibinfo{volume}{113}},
  \bibinfo{pages}{252301} (\bibinfo{year}{2014}). 

\bibitem[{\citenamefont{Zhang}(1998)}]{Zhang:1997ej}
\bibinfo{author}{\bibfnamefont{B.}~\bibnamefont{Zhang}},
  \bibinfo{journal}{Comput.Phys.Commun.} \textbf{\bibinfo{volume}{109}},
  \bibinfo{pages}{193} (\bibinfo{year}{1998}). 

\bibitem[{\citenamefont{Molnar and Gyulassy}(2000)}]{Molnar:2000jh}
\bibinfo{author}{\bibfnamefont{D.}~\bibnamefont{Molnar}} \bibnamefont{and}
  \bibinfo{author}{\bibfnamefont{M.}~\bibnamefont{Gyulassy}},
  \bibinfo{journal}{Phys.Rev.} \textbf{\bibinfo{volume}{C62}},
  \bibinfo{pages}{054907} (\bibinfo{year}{2000}), \eprint{nucl-th/0005051}.

\bibitem[{\citenamefont{Ollitrault}(1993)}]{Ollitrault:1993ba}
\bibinfo{author}{\bibfnamefont{J.-Y.} \bibnamefont{Ollitrault}},
  \bibinfo{journal}{Phys.Rev.} \textbf{\bibinfo{volume}{D48}},
  \bibinfo{pages}{1132} (\bibinfo{year}{1993}). 

\bibitem[{\citenamefont{Voloshin and Zhang}(1996)}]{Voloshin:1994mz}
\bibinfo{author}{\bibfnamefont{S.}~\bibnamefont{Voloshin}} \bibnamefont{and}
  \bibinfo{author}{\bibfnamefont{Y.}~\bibnamefont{Zhang}},
  \bibinfo{journal}{Z.Phys.} \textbf{\bibinfo{volume}{C70}},
  \bibinfo{pages}{665} (\bibinfo{year}{1996}). 

\bibitem[{\citenamefont{Molnar}(2005)}]{Molnar:2005hb}
\bibinfo{author}{\bibfnamefont{D.}~\bibnamefont{Molnar}}
  (\bibinfo{year}{2005}), \eprint{nucl-th/0503051}.
%
\bibinfo{author}{\bibfnamefont{N.}~\bibnamefont{Borghini}} \bibnamefont{and}
  \bibinfo{author}{\bibfnamefont{C.}~\bibnamefont{Gombeaud}},
  \bibinfo{journal}{Eur.Phys.J.} \textbf{\bibinfo{volume}{C71}},
  \bibinfo{pages}{1612} (\bibinfo{year}{2011}), \eprint{1012.0899}.

\bibitem[{\citenamefont{Heiselberg and Levy}(1999)}]{Heiselberg:1998es}
\bibinfo{author}{\bibfnamefont{H.}~\bibnamefont{Heiselberg}} \bibnamefont{and}
  \bibinfo{author}{\bibfnamefont{A.-M.} \bibnamefont{Levy}},
  \bibinfo{journal}{Phys.Rev.} \textbf{\bibinfo{volume}{C59}},
  \bibinfo{pages}{2716} (\bibinfo{year}{1999}); 
%
\bibinfo{author}{\bibfnamefont{S.}~\bibnamefont{Voloshin}} \bibnamefont{and}
  \bibinfo{author}{\bibfnamefont{A.~M.} \bibnamefont{Poskanzer}},
  \bibinfo{journal}{Phys.Lett.} \textbf{\bibinfo{volume}{B474}},
  \bibinfo{pages}{27} (\bibinfo{year}{2000}). 
%
\bibitem[{\citenamefont{Kolb et~al.}(2001)\citenamefont{Kolb, Huovinen, Heinz, and Heiselberg}}]{Kolb:2000fha}
\bibinfo{author}{\bibfnamefont{P.~F.} \bibnamefont{Kolb}},
  \bibinfo{author}{\bibfnamefont{P.}~\bibnamefont{Huovinen}},
  \bibinfo{author}{\bibfnamefont{U.~W.} \bibnamefont{Heinz}}, \bibnamefont{and}
  \bibinfo{author}{\bibfnamefont{H.}~\bibnamefont{Heiselberg}},
  \bibinfo{journal}{Phys. Lett.} \textbf{\bibinfo{volume}{B500}},
  \bibinfo{pages}{232} (\bibinfo{year}{2001}). 

\bibitem[{\citenamefont{Wang and Gyulassy}(1992)}]{Wang:1991xy}
\bibinfo{author}{\bibfnamefont{X.-N.} \bibnamefont{Wang}} \bibnamefont{and}
  \bibinfo{author}{\bibfnamefont{M.}~\bibnamefont{Gyulassy}},
  \bibinfo{journal}{Phys.Rev.Lett.} \textbf{\bibinfo{volume}{68}},
  \bibinfo{pages}{1480}, (\bibinfo{year}{1992}).

\bibitem[{\citenamefont{Shuryak}(2002)}]{Shuryak:2001me}
\bibinfo{author}{\bibfnamefont{E.V.}~\bibnamefont{Shuryak}},
  \bibinfo{journal}{Phys.Rev.} \textbf{\bibinfo{volume}{C66}},
  \bibinfo{pages}{027902} (\bibinfo{year}{2002}). 

\bibitem[{\citenamefont{Kovtun et~al.}(2005)\citenamefont{Kovtun, Son, and
  Starinets}}]{Kovtun:2004de}
\bibinfo{author}{\bibfnamefont{P.}~\bibnamefont{Kovtun}},
  \bibinfo{author}{\bibfnamefont{D.}~\bibnamefont{Son}}, \bibnamefont{and}
  \bibinfo{author}{\bibfnamefont{A.}~\bibnamefont{Starinets}},
  \bibinfo{journal}{Phys.Rev.Lett.} \textbf{\bibinfo{volume}{94}},
  \bibinfo{pages}{111601} (\bibinfo{year}{2005}), \eprint{hep-th/0405231}.

\bibitem[{\citenamefont{Cooper and Frye}(1974)}]{Cooper:1974mv}
\bibinfo{author}{\bibfnamefont{F.}~\bibnamefont{Cooper}} \bibnamefont{and}
  \bibinfo{author}{\bibfnamefont{G.}~\bibnamefont{Frye}},
  \bibinfo{journal}{Phys.Rev.} \textbf{\bibinfo{volume}{D10}},
  \bibinfo{pages}{186} (\bibinfo{year}{1974}).

\bibitem[{\citenamefont{Grassi et~al.}(1995)\citenamefont{Grassi, Hama, and
  Kodama}}]{Grassi:1994nf}
\bibinfo{author}{\bibfnamefont{F.}~\bibnamefont{Grassi}},
  \bibinfo{author}{\bibfnamefont{Y.}~\bibnamefont{Hama}}, \bibnamefont{and}
  \bibinfo{author}{\bibfnamefont{T.}~\bibnamefont{Kodama}},
  \bibinfo{journal}{Phys.Lett.} \textbf{\bibinfo{volume}{B355}},
  \bibinfo{pages}{9} (\bibinfo{year}{1995});
%
\bibinfo{author}{\bibfnamefont{O.}~\bibnamefont{Socolowski}},
  \bibinfo{author}{\bibfnamefont{F.}~\bibnamefont{Grassi}},
  \bibinfo{author}{\bibfnamefont{Y.}~\bibnamefont{Hama}}, \bibnamefont{and}
  \bibinfo{author}{\bibfnamefont{T.}~\bibnamefont{Kodama}},
  \bibinfo{journal}{Phys.Rev.Lett.} \textbf{\bibinfo{volume}{93}},
  \bibinfo{pages}{182301} (\bibinfo{year}{2004}). 
\end{thebibliography}

\end{document}